\documentclass[aps, prb, reprint, amsmath, amssymb,
 aps, superscriptaddress,floatfix]{revtex4-1}
\usepackage{braket}
\usepackage{graphicx}
\usepackage{hyperref}
\usepackage[caption=false]{subfig}
\usepackage{comment} 
\usepackage{amsmath} 
\usepackage{amssymb}
\usepackage{wrapfig}
\usepackage{bbm}
\usepackage{url,hyperref} 
\usepackage{color}
\begin{document}

\newcommand{\mytitle}{Signatures of the Majorana spin in electrical transport through a Majorana nanowire}
\newcommand{\I}{\mathbbm{1}}
\newcommand{\T}{\mathcal{T}}
\newcommand{\U}{\mathcal{U}}
\newcommand{\N}{\mathcal{N}}
\newcommand{\hc}{\text{H.c.}}
\newcommand{\vlambda}{\vec{\lambda}}
\newcommand{\alphabar}{{\bar{\alpha}}}
\newcommand{\MathInclude}[1]{\medskip\fbox{\includegraphics[width=165mm]{Mathematica/#1}}\medskip}

\newcommand{\dd}{\mathrm{d}\,} 
\newcommand{\sign}{\mathrm{sign}\,}
\newcommand{\tr}{\mathrm{tr}\,}
\newcommand{\adj}{\mathrm{adj}\,}
\newcommand{\diag}{\mathrm{diag}\,} 
\newcommand{\const}{\mathrm{const}\,} 
\newcommand{\figref}[1]{\figurename~\ref{#1}} 
\newcommand{\sref}[1]{Section~\ref{#1}} 

\title{\mytitle}
\author{Alexander Schuray}
\affiliation{Institut f\"ur Mathematische Physik, Technische Universit\"at
        Braunschweig, D-38106 Braunschweig, Germany}
        \author{Manuel Rammler}
\affiliation{Institut f\"ur Mathematische Physik, Technische Universit\"at
        Braunschweig, D-38106 Braunschweig, Germany}
   \author{Patrik Recher}
    \affiliation{Institut f\"ur Mathematische Physik, Technische Universit\"at
        Braunschweig, D-38106 Braunschweig, Germany}
    \affiliation{Laboratory for Emerging Nanometrology Braunschweig, D-38106 
        Braunschweig, Germany}
\date{\today}

\begin{abstract}
In this paper, we investigate the transport properties of spinful electrons tunnel coupled to a finite-length Majorana nanowire on one end which is further tunnel coupled to a quantum dot (QD) at the other end. Using a full counting statistics approach, we show that Andreev reflection can happen in two separate channels that can be associated with the two spin states of the tunneling electrons. In a low-energy model for the nanowire that is represented by two overlapping Majorana bound states (MBSs) localized at the ends of the wire, analytical formulas for conductance and noise reveal their crucial dependence on the spin-canting angle difference of the two MBSs in the absence of the QD if the spinful lead couples to both MBSs. We further investigate the influence of a finite temperature on the observation of the coupling to both MBSs. In the presence of the QD, the interference of different tunneling paths gives rise to Fano resonances and the symmetry of those provide decisive information about the coupling to both MBSs. We contrast the low-energy model with a tight-binding model of the Majorana nanowire  and treat the Coulomb interaction on the QD with a self-consistent mean field approach. Using the scattering matrix approach, we thereby extend the transport results obtained in the low-energy model including also higher excited states in the nanowire. 
\end{abstract} 
\maketitle
\section{Introduction}
\label{sec:Intro}
Since A. Kitaev showed that a one-dimensional spinless $p$-wave superconductor can host Majorana bound states (MBSs) at its boundaries~\cite{Kitaev2001} a tremendous amount of research activities focused on the creation, detection, and manipulation of MBSs~\cite{Alicea2012,Beenakker2013,Sarma2015,Aguado2017,Lutchyn2018,Zhang2019,Beenakker2019,Prada2019,Schuray2020}. These quasiparticles are not only interesting because of their fundamental property of being their own antiparticles~\cite{Majorana1937} but also because of their non-Abelian anyonic exchange statistics~\cite{Ivanov2001,Stern2004, Nayak2008}. The latter makes them particularly interesting for fault tolerant topological quantum computation schemes~\cite{Kitaev2003,Nayak2008,Pachos2012,Lahtinen2017, Beenakker2019}.

In general, $p$-wave superconductivity needs to be designed using hybrid structures. One of the first suggested experimental realizations of these exotic superconductors is a semiconducting nanowire with Rashba spin orbit coupling and proximity induced $s$-wave superconductivity where an applied Zeeman field drives a topological phase transition~\cite{Lutchyn2010,Oreg2010,Aguado2017,Prada2019}. By now many different schemes to create MBSs have been proposed and realized, for example magnetic adatoms on superconductor surfaces~\cite{Choy2011,Nadj-Perge2013,Klinovaja2013,Braunecker2013,Vazifeh2013,Pientka2013,Nadj-Perge2014,Pawlak2016,Jeon2017,Ruby2017}, helical edge or hinge modes with competing superconducting and magnetic gap opening mechanisms~\cite{Fu2008,Langbehn2017,Schindler2018,Jack2019,Novik2019,Fleckenstein2020}, or topological Josephson junctions~\cite{Fu2008,Grosfeld2011,Potter2013,Park2015,Pientka2017,Hell2017,Ren2019,Scharf2019}. However, most experimental reports to date on the existence of MBSs focus on the nanowire setup~\cite{Mourik2012,Rokhinson2012,Deng2012,Das2012,Churchill2013,Finck2013,Albrecht2016,Deng2016,Chen2017,Suominen2017,Nichele2017,Guel2018,Zhang2018,Sestoft2018,Deng2018,Laroche2019}.

In order to establish the existence of MBSs different transport signatures have been suggested. First of all, tunneling into an isolated MBS at very low temperatures leads to a robust quantized zero bias differential conductance of $2e^2/h$~\cite{Law2009,Flensberg2010} and recent experimental data showed that value in tunneling experiments~\cite{Nichele2017,Zhang2018}. Other suggestions include the fractional $4\pi$-periodic Josephson effect in a topological Josephson junction~\cite{Kitaev2001,Kwon2004,Fu2009b,San-Jose2012,Dominguez2012,Beenakker2013,Virtanen2013,Houzet2013,Crepin2014,Lee2014,Kane2015,Peng2016,Kuzmanovski2016,Pico-Cortes2017,Dominguez2017,Klees2017,Cayao2017,Sticlet2018,Frombach2019} or the change from a $2e$ to an $e$ periodicity in Coulomb blockade resonances in the case of a floating nanowire where charging effects become relevant~\cite{Fu2010,Zazunov2011,Hutzen2012,vanHeck2016,Lutchyn2017,Ekstrom2019}. Both effects have been observed experimentally~\cite{Rokhinson2012,Albrecht2016,Wiedenmann2016,Bocquillon2016,Deacon2017,Laroche2019}. Even though the evidence for the existence of MBSs is ever growing an unambiguous proof remains elusive.

To gain additional Majorana signatures it is also possible to couple the MBSs to a quantum dot (QD)~\cite{Liu2011,Leijnse2011,Vernek2014}. In setups containing MBSs and QDs Fano resonances (FRs) for which a resonant path interferes with a continuous path~\cite{Miroshnichenko2010} can emerge in the differential conductance. These FRs can either manifest themselves as a function of applied bias voltage~\cite{Dessotti2014,Xia2015,Xiong2016,Baranski2016,Wang2018b,Ricco2018} when the QD is directly coupled by a lead, as a function of flux through a loop with MBSs~\cite{Ueda2014,Jiang2015,Zeng2016} or as a function of dot level energy~\cite{Schuray2017}. Also, recent experiments showed that it is possible to couple a QD to a Majorana nanowire~\cite{Deng2016,Deng2018}.

However, these experiments showed a hybridization between dot and low-energy in-gap states of the Majorana nanowire that was not compatible with coupling to a single MBS, but could be explained with a coupling to both MBSs~\cite{Schuray2017,Clarke2017,Prada2017} when the Majorana wave functions reach the other end of the wire. This hybridization can also be used to define a quality factor or degree of locality of the two MBSs~\cite{Clarke2017,Prada2017,Penaranda2018}. Due to the interplay between Rashba spin orbit coupling and Zeeman field, there is no homogenous spin quantization axis along the nanowire. This nontrivial spin structure is transferred to the Majorana spinor wave function~\cite{Sticlet2012,Prada2017,Ptok2017,Serina2018,Milz2019}, so that the spins of the two MBSs at the same position can point in different directions, which can influence the transport properties~\cite{Schuray2018}.

Recently, it was pointed out that the signatures of MBSs can be mimicked by trivial Andreev bound states with partially separated Majorana components that arise at the interface of an N-S junction~\cite{SanJose2016,Liu2017,Reeg2018,Vuik2018,Moore2018,Fleckenstein2018,Liu2018,Avila2019,Stanescu2019,Zhang2019,Pan2019,Oladunjoye2019}. It is therefore of utmost importance to find irrevocable signatures of MBSs. Thus, we propose in this work that the spin-canting angles of the Majorana components from the two ends of the nanowire is another tool that can be used to distinguish between topological MBSs and these trivial states that are also dubbed nontopological MBSs~\cite{Prada2019}. 

In this paper, we consider two different scenarios. First, we consider a finite-length Majorana nanowire tunnel coupled to a spinful lead. Here, we put special emphasis on the finite size of the nanowire which allows us to probe both MBSs wave functions via the coupling to the lead. Second, we consider a Majorana nanowire tunnel coupled to a lead on one side and a QD on the other side. We include the finite length of the nanowire and thus allow for a tunnel coupling of the lead and the dot to both MBSs. Differently from our previous work~\cite{Schuray2017}, we put emphasis on the spin degree of freedom and find that the spin-canting angle of the MBSs have profound consequences on the transport properties and that the spin degree of freedom in the lead and the QD can not be omitted for realistic system parameters.

We use full counting statistics (FCS) together with an effective low-energy model to show that the only processes contributing to transport via the MBSs are Andreev reflections via two different electronic channels in the lead. In the absence of a tunnel coupling between nanowire and QD, we show analytically that the differential conductance is a function of the spin-canting angle difference of the two MBSs at the junction with the lead and that one channel is blocked if both spins point in the same direction or the coupling to the distant MBS vanishes. Moreover, we show the emergence of two pairs of Fano resonances as a function of dot level energy in the case where the dot is tunnel coupled to the nanowire. We find that the symmetry relation within each pair of resonances unveils if the coupling between dot and wire consists of coupling to one MBS only or to both of them. In the former case, the Hamiltonian obeys an approximate electron-hole symmetry with respect to a reversal of the QD level energy which is absent in the latter case.

The paper is organized as follows. In Sec.~\ref{sec:model} we introduce the model that underlies our calculations. We calculate the cumulant generating function, the main entity for the FCS, using the Keldysh Green's function formalism from which we extract all transport properties in Sec.~\ref{sec:FCS}. After we have established our model and method we consider the special case of a vanishing tunnel coupling between MBSs and QD in Sec.~\ref{sec:FCS}~A, before discussing our results with a finite coupling between MBSs and QD in Sec.~\ref{sec:FCS}~B. To underline our findings we also calculate the differential conductance numerically using a discretized Rashba wire model which allows also the inclusion of excited states in Sec.~\ref{sec:numerical}. We again consider first the system without QD in Sec.~\ref{sec:numerical}~A and the system with QD in Sec.~\ref{sec:numerical}~B where we treat Coulomb interactions on the QD with a self-consistent mean field approximation.

\section{Model}
\label{sec:model}
\begin{figure}
	\centering
\includegraphics[width=\columnwidth]{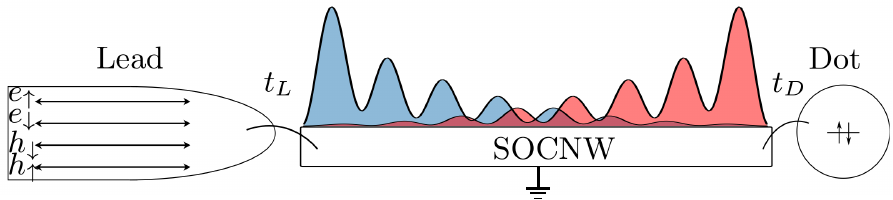}
	\caption{Schematic representation of the setup under consideration. The normal spinful lead contains two electron (up, down) and two hole (up, down) channels. We assume a pointlike tunneling from the lead ($t_L$) and dot ($t_D$) to the corresponding ends of the nanowire. Blue (red) are the calculated Majorana wave functions of the left (right) ends of the grounded spin orbit coupled Majorana nanowire (SOCNW).}
	\label{fig:setup}
\end{figure}
We calculate the electronic transport properties of a normal spinful lead and a nanowire-superconductor-QD hybrid structure. A schematic representation of the setup under consideration is shown in Fig.~\ref{fig:setup}.
The nanowire in proximity to a grounded s-wave superconductor is described using a noninteracting effective mass approximation and the Bogoliubov-de Gennes formalism. The Hamiltonian is
\begin{equation}
	H_{NW}=\frac{1}{2}\int_0^L \Psi^\dagger(x) \mathcal{H}_{BdG}^{NW} \Psi(x) dx, 
	\label{NWBdG}
\end{equation}
where $L$ is the length of the nanowire and $H_{BdG}^{NW}$ is presented in the Nambu basis with $\Psi(x)=\left[\psi_\uparrow(x),\psi_\downarrow(x),\psi_\downarrow^\dagger(x),-\psi_\uparrow^\dagger(x)\right]^T$, and
 \begin{align}
	\mathcal{H}_{BdG}^{NW}=&\left[\left(-\frac{\hbar^2}{2m^*}\partial_x^2 -\mu\right)-i \alpha \partial_x \sigma_y\right]\tau_z\notag\\
	&+V_Z\sigma_z+\Delta\tau_x.	
\end{align}
Here, $m^*$ is the effective electron mass, $\mu$ is the chemical potential, $\alpha$ is the Rashba parameter, $V_Z$ is the Zeeman energy and $\Delta$ is the proximity induced $s$-wave pairing. The Pauli matrices $\sigma_i$ and $\tau_i$ act in the spin and particle-hole space, respectively. The topological nontrivial phase with emerging MBSs is present for $V_{Z}>\sqrt{\Delta^2+\mu^2}$~\cite{Lutchyn2010,Oreg2010}.
The low-energy sector of the nanowire in the topologically nontrivial regimes is governed by the two MBSs forming at the ends of the nanowire.

The normal lead is spin degenerate (metallic regime), an assumption which has been used also in similar setups~\cite{Leijnse2011,Haim2015,Baranski2016,Stenger2017,Avila2019}. We further linearize the spectrum around the Fermi energy, which is a valid assumption for a metallic lead since the bias voltage and temperature of interest are small compared to the Fermi energy. Its Hamiltonian is therefore given by
\begin{equation}
	H_L=-i\hbar v_F\sum_\sigma\int dx c^{\dagger}_\sigma(x)\partial_xc_\sigma(x),
	\label{HL}
\end{equation}
where $c^\dagger_\sigma(x)$ creates an electron with spin $\sigma$ at position $x$ and $v_F$ is the Fermi velocity in the lead. 

The QD is modeled using a single electronic level with energy $\varepsilon_D$ which can be empty, occupied by one electron (spin up or spin down) or doubly occupied (spin singlet). Due to the Coulomb interaction between the electrons the double occupancy results in an additional energy cost $U$. $\varepsilon_D$ is experimentally unable by a gate voltage. In addition, we include the same Zeeman field as in the nanowire, because in current experiments~\cite{Deng2016,Deng2018} the fabricated QD is made out of the same material as the nanowire. The QD Hamiltonian is
\begin{equation}
	H_D=\sum_{\sigma,\sigma'}d^{\dagger}_{\sigma}\big[\varepsilon_D\sigma_0+V_Z\sigma_z\big]_{\sigma\sigma'}d_{\sigma'}+Un_{\uparrow}n_{\downarrow}.
	\label{HD}
\end{equation}
We note that possible effects of the spin orbit coupling are neglected which is appropriate for small QDs \cite{Hanson2007,Hoffman2017}.
We use a mean field approximation for the Coulomb interaction~\cite{Prada2017} which results in 
\begin{equation}
	Un_{\uparrow}n_{\downarrow}\approx U\left( n_\uparrow\braket{n_\downarrow}+\braket{n_\uparrow}n_\downarrow-\braket{n_\uparrow}\braket{n_\downarrow} \right).
	\label{CoulombH}
\end{equation}
 It is important to note that quadratic fluctuations in the dot occupation number are neglected in this approximation. In addition, we note that the mean field approximation would not be able to capture Kondo physics~\cite{MartinRodero2011}. However, Kondo correlations are suppressed by the applied Zeeman field (i.e., if $\Delta_Z$ is larger than the Kondo temperature).

The coupling between the nanowire and the lead and the QD is described with the tunneling Hamiltonian
\begin{equation}
	H_T=\sum_\sigma t_{L}c^{\dagger}_\sigma(0)\psi_\sigma(0)+t_{D}d^\dagger_\sigma\psi_\sigma(L) + \hc,
	\label{HT}
\end{equation}
where we assume a pointlike tunneling between the nanowire ends and lead and dot, respectively.
For the case of an extended barrier between nanowire and QD, see Ref.~\onlinecite{Hoffman2017}.
The Hamiltonian of the complete system is therefore given by
\begin{equation}
	H=H_{NW}+H_L+H_D+H_T.
	\label{Hamiltonian}
\end{equation}
In our calculation we do not aim at a quantitative agreement with recent experiments, but we use realistic microscopic parameters to underline the relevance of our findings for current and future experimental efforts. If not explicitly stated otherwise we will use the parameters $m^*=0.015m_e$ where $m_e$ is the electron rest mass, $\alpha=20$ meVnm, and $\Delta=0.5$ meV (cf. Ref.~\onlinecite{Bommer2019}).

\section{Low-energy model and full counting statistics}
\label{sec:FCS}
To calculate the transport properties of our proposed system we resort to FCS. Because we are interested in the signatures from the MBSs we consider an effective low-energy Hamiltonian for the nanowire. For simplicity we set $\hbar=e=1$. We will restore the units for the main transport results. The low-energy sector of the Hilbert space of the nanowire is composed of the two MBSs and the effective Hamiltonian is 
\begin{equation}
H_{Eff}=i\varepsilon\gamma_1\gamma_2,
\label{Heff}
\end{equation}
where $\gamma_i$ is the Hermitian creation operator for the $i$-th MBS and $\varepsilon$ is the hybridization energy of these two MBSs. The Majorana operators satisfy the anticommutator relation $\{\gamma_i,\gamma_j\}=2\delta_{ij}$. This Hamiltonian is diagonalized by the nonlocal fermion $\eta=(\gamma_1+i\gamma_2)/2$ which can be either empty or occupied. Also, the annihilation operator can be expressed in this approximation using the MBSs operators $\psi_\sigma(x)=\Lambda_{1\sigma}(x)\gamma_1+\Lambda_{2\sigma}(x)\gamma_2$ where $\Lambda_{i\sigma}(x)$ is the electronic part of the spinor wave function of the $i$-th MBS. In this model, the MBS wave function has no spin-$y$ component, so that we can use the parametrization
\begin{equation}
\label{spinorMBS}
 \left(\begin{matrix}
        \Lambda_{i\uparrow}(x)\\
        \Lambda_{i\downarrow}(x)
       \end{matrix}
\right)= \kappa_i(x)\left(\begin{matrix}
        \cos\left( \frac{\Theta_i(x)}{2} \right)\\
        \sin\left( \frac{\Theta_i(x)}{2} \right)
       \end{matrix}
\right),
\end{equation}
where $\Theta_i(x)$ is the spin-canting angle of the $i$-th MBS at postion $x$ and $\kappa_i(x)$ is the spatial profile of the wave function. 

With this parametrization and the decomposition of the field operators $\psi_{\sigma}(x)$ into the Majorana operators $\gamma_{1,2}$ we can rewrite the tunneling amplitudes in Eq.~(\ref{HT}) as
\begin{align}
t_L\Lambda_{i\uparrow}(0)&=t_i\cos\left( \frac{\Theta_i(0)}{2} \right)\equiv t_{Li\uparrow},\\
t_L\Lambda_{i\downarrow}(0)&=t_i\sin\left( \frac{\Theta_i(0)}{2} \right)\equiv t_{Li\downarrow},\notag\\
t_D\Lambda_{i\uparrow}(L)&=t_{Di}\cos\left( \frac{\Theta_i(L)}{2} \right)\equiv t_{Di\uparrow},\notag\\
t_D\Lambda_{i\downarrow}(L)&=t_{Di}\sin\left( \frac{\Theta_i(L)}{2} \right)\equiv t_{Di\downarrow}\notag.
\label{tparams}
\end{align}
The angles $\Theta_i(x)$ are the spin-canting angles of the two MBSs at position $x$ in the wire. Because we base our effective Hamiltonian upon Eq.~(\ref{NWBdG}), we find for the spin-canting angles $\Theta_1(0)=\Theta_2(L)=\Theta_1$ and $\Theta_2(0)=-\Theta_1(L)=\Theta_2$. The spin-canting angles are functions of all microscopic parameters of the nanowire and we refer an interested reader to Refs.~\onlinecite{Prada2017,Schuray2018} for a more in depth analysis of the spin-canting angles. Moreover, we can transform the creation (annihilation) operators of the QD into a Majorana operator basis
\begin{align}
d^\dagger_{\uparrow}=&\frac{1}{2} \left(\gamma_3+i\gamma_4\right)\\
d^\dagger_{\downarrow}=& \frac{1}{2}\left( \gamma_5+i\gamma_6 \right).\notag
\end{align}
We then can rewrite $H$ in the low-energy sector as
\begin{equation}
H=H_M'+H_T'+H_L,
\label{Hfull}
\end{equation}
with
\begin{align}
H_M'&=\frac{i}{2}\sum_{\mu\nu}A_{\mu\nu}\gamma_\mu\gamma_\nu,
\label{Hs}\\
H_T'&=\sum_{i\sigma}t_{Li\sigma}c^{\dagger}_\sigma(0)\gamma_i+ \hc,\notag
\end{align}
where the matrix $A$ contains the tunneling between the dot and the MBSs as well as the Majorana hybridization energy and the single particle energies of the two spin states of the QD which are shifted by the Coulomb interaction in the mean field approximation.

The cumulant generating function (CGF) can be expressed in the Levitov-Lesovik form~\cite{Nazarov2003,Levitov2004,Weithofer2014} (a derivation can be found in Appendix~\ref{sec:A})
\begin{equation}
\ln\chi(\lambda)=\frac{\mathcal{T}}{2}\int\frac{d\omega}{2\pi}\ln\left[\frac{\det\left([D^{\lambda}]^{-1}(\omega)\right)}{\det\left([D^{\lambda=0}]^{-1}(\omega)\right)}\right],
\label{LevitovII}
\end{equation}
where $\mathcal{T}$ is a long measuring time and the inverse full Majorana Green's function $[D^{\lambda}]^{-1}(\omega)=[D^{(0)}(\omega)]^{-1}-\Sigma^{\lambda}(\omega)$ is a $12\times 12$ matrix ($2$ for the MBSs, $4$ for the electron and hole degrees of freedom on the dot and a factor 2 because of the Keldysh formalism). Here, $D^{(0)}(\omega)$ is the unperturbed Majorana Green's function, the Fourier transform of $[D^{(0)}(t,t')]_{\alpha\beta}=-i\braket{T_\mathcal{C}\gamma_\alpha(t)\gamma_\beta(t')}$ with $T_\mathcal{C}$ being the time-ordering operator on the Keldysh contour, and $\Sigma^{\lambda}(\omega)$ is the Fourier transform of the self energy containing the counting field
\begin{align}
\Sigma^{\lambda}_{\alpha\beta}(t,t')=&\sum_\sigma\bigg[-t_{L\alpha\sigma}t^*_{L\beta\sigma}e^{-i\frac{\lambda(t)-\lambda(t')}{2}}G_\sigma(t,t')\\
&+t_{L\beta\sigma}t^*_{L\alpha\sigma}e^{i\frac{\lambda(t)-\lambda(t')}{2}}G_\sigma(t',t)\bigg],\notag
\end{align} 
where $G_\sigma(t',t)=G_\sigma(x'=0,x=0,t',t)=-i\braket{T_{\mathcal{C}}c_\sigma(x'=0,t')c_\sigma^\dagger(x=0,t)}$ is the unperturbed lead boundary Green's function for spin $\sigma$ and $t_{L\alpha\sigma}=0$ for $\alpha>2$. The detailed calculations of these Green's functions can be found in Appendix~\ref{sec:B}.
For general temperatures, we find the CGF
\begin{align}
\ln\chi(\lambda)=&\frac{\mathcal{T}}{2}\sum_{i=\pm}\int\frac{d\omega}{2\pi}\ln\bigg[1+p_i\left( e^{-2i\lambda}-1 \right)n(\omega)n(-\omega)\notag\\
&+p_i\left(e^{2i\lambda}-1\right)\left( n(\omega)-1 \right)\left( n(-\omega)-1 \right)\bigg],
\label{lnchiT}
\end{align}
where $n(\omega)=\frac{1}{1+e^{\beta(\omega-V)}}$ is the Fermi function in the lead with $\beta=1/k_BT$ the inverse thermal energy and $V$ the bias voltage between the lead and the grounded superconductor. It shows that Andreev reflection is the only transport process which contributes to the current.
The probability for an Andreev reflection in a given channel at energy $\omega$ is given by $p_i(\omega)$. At zero temperature, the CGF corresponds to a generalized binomial distribution [see Eq.~(\ref{cgfwd})].
The main difference from previous works considering the FCS of a lead coupled to a MBS system~\cite{Weithofer2014,Schuray2017} is that in the spinful case there are two channels in which Andreev reflections are possible. These two channels originate from the two spin channels for the electrons.

The average current and the symmetrized zero-frequency noise can easily be calculated from the CGF by taking the first or second derivative with respect to the counting field, respectively
\begin{align}
I&=\frac{i}{\mathcal{T}}\frac{d}{d\lambda}\ln\left( \chi(\lambda) \right)|_{\lambda=0}\\
P&=\frac{-1}{\mathcal{T}}\frac{d^2}{d\lambda^2}\ln\left( \chi(\lambda) \right)|_{\lambda=0}.
\label{IandP}
\end{align}
\begin{figure*} 
	\centering
\includegraphics[width=\textwidth]{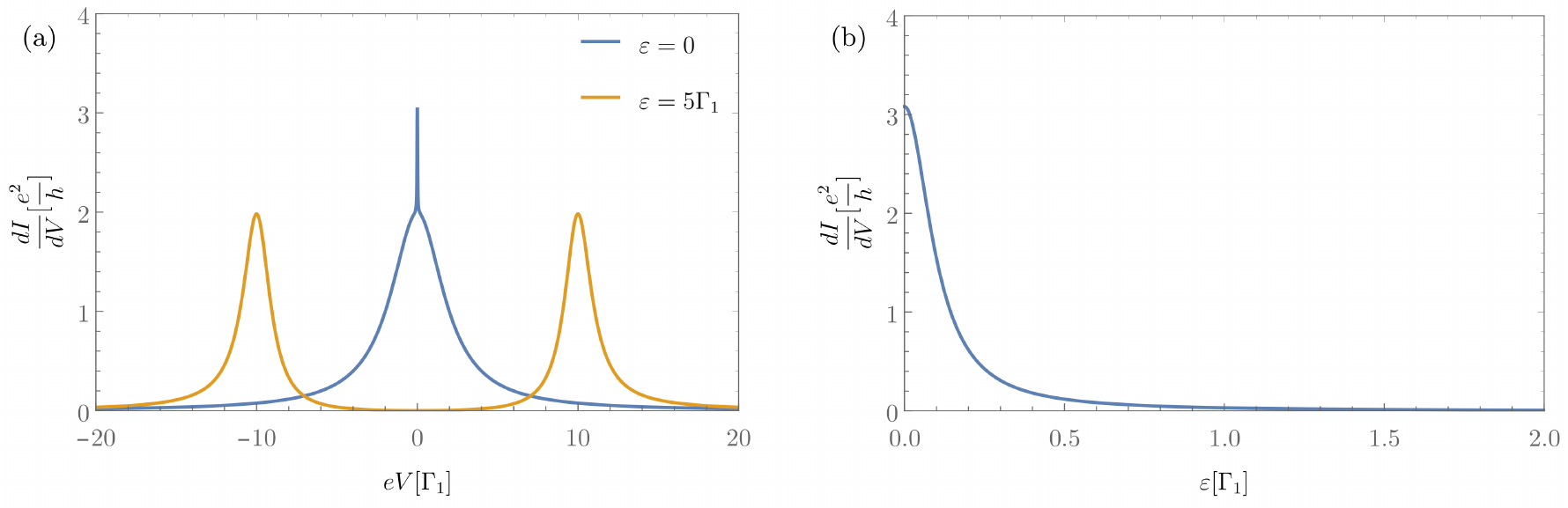}
	\caption{Differential conductance in the low-energy model in the setup without QD with finite coupling to both MBS. (a) Differential conductance as a function of bias energy with zero Majorana splitting energy (blue) and finite splitting energy $\varepsilon=5\Gamma_1$ (yellow). For $\varepsilon=0$, the differential conductance has the shape of a sum of two Lorentzians with two different widths and deviates strongly from the expected value at the resonance in the case of a coupling to one MBS. Whereas in the case of large splitting energy $\varepsilon\gg\Gamma_2$, such deviations at the resonance are very small.  (b) Zero bias differential conductance vs Majorana splitting energy. The differential conductance at zero bias is not quantized for $\varepsilon=0$ and does not vanish for finite splitting energies. The other parameters are  $\Gamma_2=0.01\Gamma_1$ and $\delta\Theta=\pi-1$.}
	\label{fig:didveff}
\end{figure*}
\subsection{A. Transport properties without quantum dot}
\label{sec:withoutdot}
\begin{figure*}
\includegraphics[width=\textwidth]{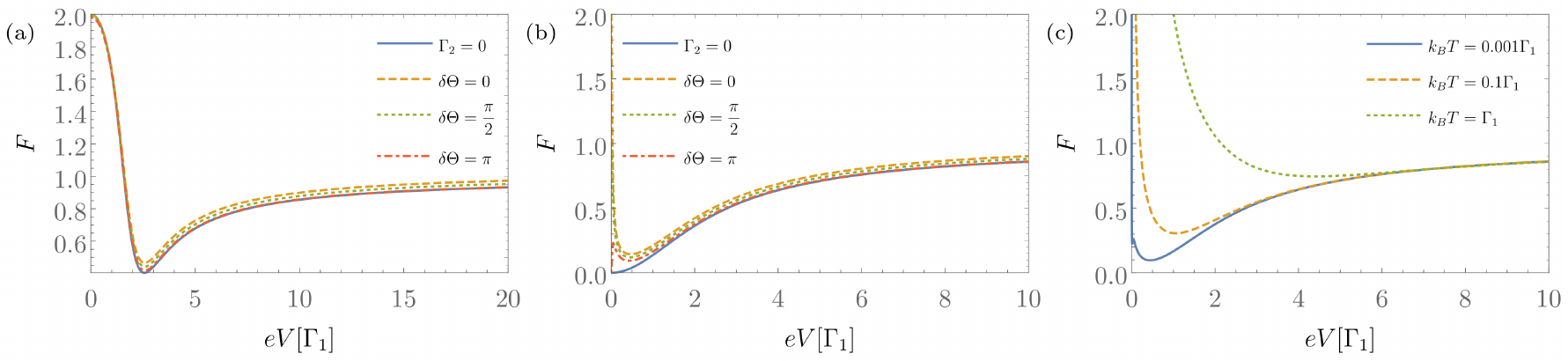}
\caption{Fano factor in the low-energy model as a function of applied bias voltage energy in the setup without QD. If not stated otherwise  $\Gamma_2=0.01\Gamma_1$. (a), (b) Fano factor  for various spin canting angle differences $\delta\Theta$ and tunneling rates to the second MBS at zero temperature. In (a) the Majorana splitting energy is finite $\varepsilon=\Gamma_1$ and in (b) it is assumed to be zero. (c) Fano factor with $\varepsilon=0$  and $\delta\Theta=\pi$ for various temperatures. }
	\label{fig:fanofac}
\end{figure*}
In this section, we want to consider the case where the couplings to the QD are set to zero ($t_D=0$). Then we are left with a spinful lead which includes a coupling to both MBSs due to the spreading of the Majorana spinor wave function along the nanowire. To the best of our knowledge, there are so far no publications using this conceptionally simple setup for an analytical analysis of the transport signatures in the spinful lead. However, the special cases of antiparallel spins~\cite{Vuik2018} or parallel spins~\cite{Ren2018} of the two MBSs in an effective model calculation have already been considered. Our findings are consistent with the results of these previous works.

At zero temperature, the CGF of this setup reads
\begin{equation}
	\ln\left(\chi(\lambda)\right)= \frac{\mathcal{T}}{2}\int_{-V}^V\frac{d\omega}{2\pi}\sum_{j=\pm}\ln\left(1+p_j(\omega)(e^{-2i\lambda}-1)\right),
	\label{cgfwd}
\end{equation}
where
\begin{widetext}\begin{footnotesize}
	\begin{align}
	p_\pm(\omega)=&2\frac{2\Gamma_1\Gamma_2(4\Gamma_1\Gamma_2+4\varepsilon^2+\omega^2)\sin^2(\frac{\delta\Theta}{2})+\omega^2(\Gamma_1-\Gamma_2)^2}{16\left(\Gamma_1\Gamma_2+\varepsilon^2\right)^2+2\left(2(\Gamma_1^2+\Gamma_2^2)-4\varepsilon^2\right)\omega^2+\omega^4}\\
	&\pm2\frac{\sqrt{(\Gamma_1-\Gamma_2)^4\omega^4+16\Gamma_1\Gamma_2\omega^2\left( (\Gamma_1+\Gamma_2)^2\varepsilon^2+(\Gamma_1-\Gamma_2)^2\omega^2 \right)\sin^2(\frac{\delta\Theta}{2})+4\Gamma_1^2\Gamma_2^2\left( (\Gamma_1-\Gamma_2)^2-4\varepsilon^2 \right)\omega^2\sin^2(\delta\Theta)}}{16\left(\Gamma_1\Gamma_2+\varepsilon^2\right)^2+2\left(2(\Gamma_1^2+\Gamma_2^2)-4\varepsilon^2\right)\omega^2+\omega^4}\notag,
	\label{ppm}
\end{align}\end{footnotesize}
\end{widetext}
with $\Gamma_i=2\pi\nu(0)|t_i|^2$ and $\delta\Theta=\Theta_1-\Theta_2$, where $\nu(0)=1/2\pi v_F$ is the density of states per spin at the Fermi level in the lead.
The differential conductance is then given by
\begin{widetext}
\begin{equation}
	\frac{dI}{dV}=\frac{2e^2}{h}\left(p_+(eV)+p_-(eV)\right)=\frac{8e^2}{h}\frac{(eV)^2(\Gamma_1-\Gamma_2)^2+2\Gamma_1\Gamma_2\sin^2(\frac{\delta\Theta}{2})(4\varepsilon^2+(eV)^2+4\Gamma_1\Gamma_2)}{(4\varepsilon^2-(eV)^2)^2+8\Gamma_1\Gamma_2(4\varepsilon^2+2\Gamma_1\Gamma_2)+4(eV)^2(\Gamma_1^2+\Gamma_2^2)}.
	\label{didvwd}
\end{equation}
\end{widetext}
At $V=0$ the differential conductance simplifies considerably and can be written as
\begin{equation}
	\frac{dI}{dV}|_{V=0}=\frac{4e^2}{h}\frac{\Gamma_1\Gamma_2}{\varepsilon^2+\Gamma_1\Gamma_2}\sin^2\left( \frac{\delta\Theta}{2} \right),	
	\label{didvv0}
\end{equation}
which describes a Lorentzian as a function of the Majorana splitting energy $\varepsilon$ with width $\sqrt{\Gamma_1\Gamma_2}$ and height $\frac{4e^2}{h}\sin^2\left( \frac{\delta\Theta}{2} \right)$. It is important to note that Eq.~(\ref{didvv0}) was derived with the assumption that either $\Gamma_2\neq0$ or $\varepsilon\neq0$.

As expected, in the case of $\Gamma_2=0$ and/or $\delta\Theta=0$, $p_-$ vanishes, because only one spin channel in the lead then couples to the MBSs for all $V$. In both cases the differential conductance is maximally $\frac{2e^2}{h}$. However, for $\Gamma_2\neq0$ and $\delta\Theta=0$ this value for the differential conductance is not even reached at the resonances.
Another interesting parameter regime is $\delta\Theta=\pi$. In this case each spin channel in the lead couples to a different MBS. The differential conductance can then reach values over $\frac{2e^2}{h}$ and is even quantized with $\frac{4e^2}{h}$ for vanishing Majorana overlap at zero bias, because then the two Andreev reflection probabilities each become a Lorentzian with width $\Gamma_i$, respectively. 

In general, depending on the parameters the differential conductance is between 0 and $\frac{4e^2}{h}$ as seen in Fig.~\ref{fig:didveff}.  This reflects the fact, as seen in the CGF, that Andreev reflection is possible in two channels (due to spin). The differential conductance at zero bias deviates from the quantized differential conductance which is one of the key signatures of tunneling into a single MBS because of the (small) coupling to the second MBS. For nearly opposite spin-canting angles and small splittings the differential conductance is approximately the sum of two Lorentzians which can be also seen in Fig.~\ref{fig:didveff}(a). The width of two Lorentzians can be vastly different. A detailed mathematical framework that analyses the emergence of the two different tunneling rates that determine the width of Lorentzians can be found in Ref.~\onlinecite{Avila2019}.

We calculate the differential noise at zero temperature to be
\begin{equation}
 \frac{dP}{dV}=\frac{4e^3}{h}\left(p_+(1-p_+)+p_-(1-p_-)\right).
\end{equation}
By measuring the differential conductance and noise both Andreev reflection probabilities can be extracted experimentally, employing the expression
\begin{equation}                                                                            
p_\pm=\frac{4h}{e^2}\frac{dI}{dV}\mp(\pm)\sqrt{\frac{4h}{e^2}\frac{dI}{dV}-\left(\frac{4h}{e^2}\frac{dI}{dV}\right)^2-\frac{8h}{e^3}\frac{dP}{dV}}.\notag                                                                                                                \end{equation}
The ambiguity (``$(\mp)$'') in this expression comes from the fact that due to the spin rotation invariance in the lead $p_+$ and $p_-$ cannot be distinguished experimentally.
We also analyze the Fano factor
\begin{equation}
	F=\frac{P}{eI},
	\label{Fanofactor}
\end{equation}
which is shown in Fig.~\ref{fig:fanofac}.

In general, the Fano factor is between $0$ and $2$. In the case of a large splitting energy $\varepsilon>\Gamma_i$ there is no qualitative difference of the Fano factor between the case of coupling to only one MBS and the case of an additional small coupling to the second MBS [Fig.~\ref{fig:fanofac} (a)]. In the case of zero energy MBSs, however, the behavior of the Fano factor is fundamentally different at low applied voltages for the case of coupling to only one MBS as compared to coupling to both MBSs as seen in Fig.~\ref{fig:fanofac}(b).

In general, we cannot find an analytical expression for the Fano factor. However, at zero bias voltage, zero temperature and finite MBS splitting energy manage to find
\begin{equation}
	F|_{V=0}=2\frac{\Gamma_1\Gamma_2\cos^2\left( \frac{\delta\Theta}{2} \right)+\varepsilon^2}{\Gamma_1\Gamma_2+\varepsilon^2}=2-\frac{h}{2e^2}\frac{dI}{dV}|_{V=0}.	
	\label{FV0}
\end{equation}
In the case of coupling to only one MBS or to only one spin direction ($\delta\Theta=0$) the Fano factor at zero bias is quantized to $2$. This Fano factor of $2$ corresponds to Cooper pairs being transferred between the lead and the superconductor. At finite temperatures, the Fano factor diverges at zero bias due to the thermal noise.
\begin{figure*}
	\includegraphics[width=\textwidth]{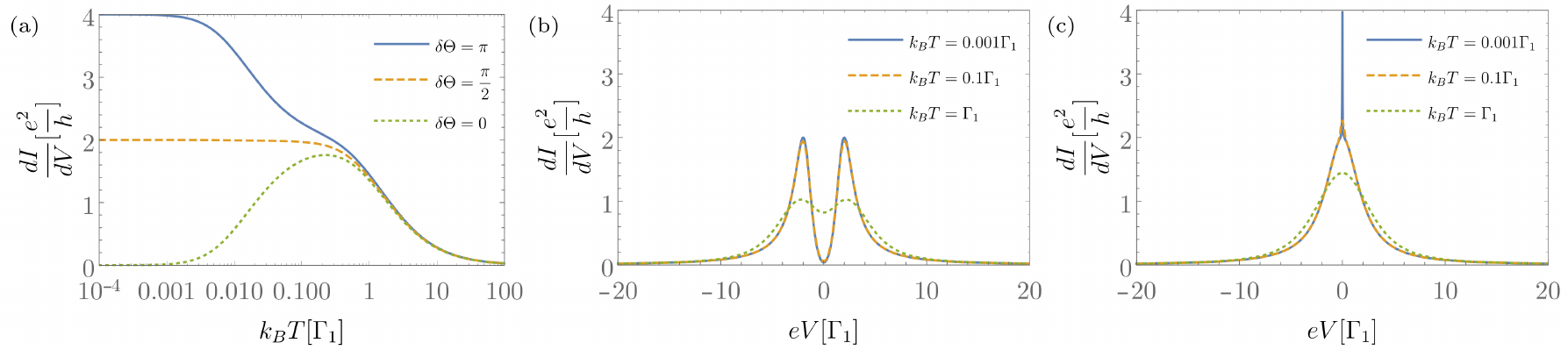}
	\caption{Differential conductance in the low-energy model in the setup without QD with finite coupling to both MBS ($\Gamma_2=0.01\Gamma_1$) at finite temperature. (a) Differential conductance at zero applied bias voltage energy ($eV=0$) for three different spin canting angle differences $\delta\Theta$ for zero Majorana splitting energy ($\varepsilon=0$). At very small temperatures ($k_{\text{B}}T\ll \Gamma_2$) the differential conductance shows plateaus. (b), (c) Differential conductance as a function of applied bias voltage energy at different finite temperatures with spin canting angle difference $\delta\Theta=\pi$. The Majorana splitting energy is finite ($\varepsilon=\Gamma_1$) in (b) and zero in (c).}
	\label{fig:tempdidv}
\end{figure*}

Finite temperatures have also an interesting influence on the differential conductance as shown in Fig.~\ref{fig:tempdidv}. In the case of zero MBS splitting and for thermal energies below all tunneling rates the differential conductance plateaus at the corresponding zero temperature value, whereas at higher temperatures ($k_BT>\Gamma_1$) the differential conductance becomes rather independent of the canting angle difference. In an intermediate temperature regime ($\Gamma_2<k_BT<\Gamma_1$) for zero energy MBSs the Lorentzian attributed to the coupling to the second MBS shrinks, while the other Lorentzian is more or less unaffected by a finite temperature as seen in Fig.~\ref{fig:tempdidv}(c).
\subsection{B. Transport properties including the quantum dot} 
\label{sec:resultdot}

First signatures of the nonlocal couplings discussed in the previous section have been seen in experiments using a QD coupled to a lead and a Majorana nanowire~\cite{Deng2016,Deng2018}. From a theoretical viewpoint the Majorana nanowire-QD setup including nonlocal couplings has been investigated spectroscopically~\cite{Prada2017,Clarke2017}. Here, we are interested in the transport signatures. In addition, we propose a setup where the QD is coupled not directly to the lead (see Fig.~\ref{fig:setup}). This is different from recent theory papers and experiments but has the advantage that the dependence on the QD level energy in the conductance probes nonlocal features of the MBSs system. In this section we focus on the low-energy transport regime of the lead-MBSs-QD system.

\begin{figure*}
\includegraphics[width=\textwidth]{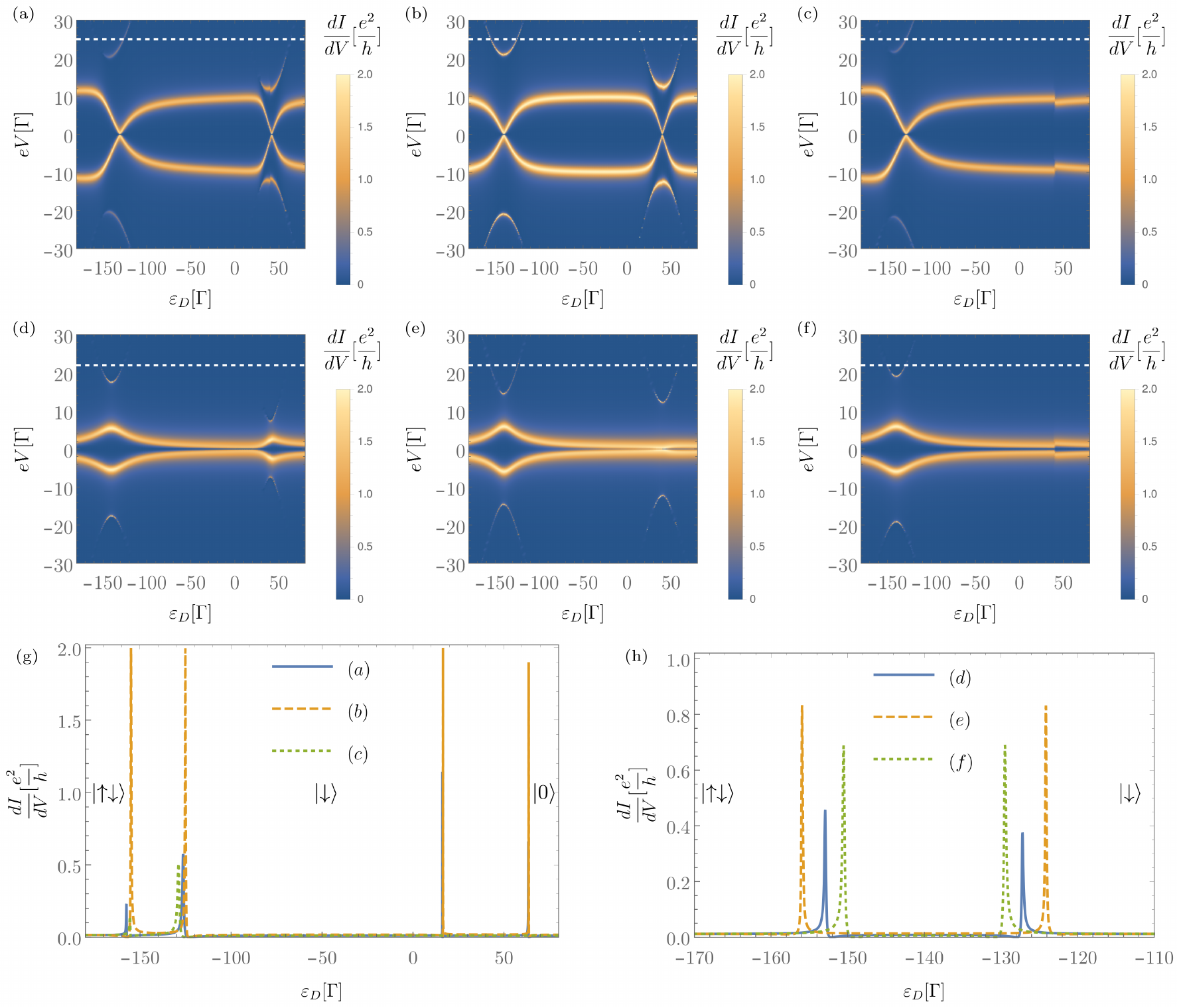}
\caption{(a)-(f) Differential conductance in the low-energy model as a function of applied bias voltage energy $eV$ and dot level energy $\varepsilon_D$ with $t_{\text{Dot}}=10\Gamma$, $U=100\Gamma$, and $V_Z=40\Gamma$. (a)-(c) show the differential conductance at finite splitting energy $\varepsilon=5\Gamma$. In (a) and (b) the spin canting angles are $\Theta_1=\Theta_2=0.8$. Here both anticrossings are visible. In (a) $\phi=0.3$. This resulting nonlocality manifests in the asymmetry  of $dI/dV$ peaks around the points where the occupancy of the dot changes. In (b) $\phi=0$ and the differential conductance shows a bow-tie shape. In (c) we consider $\phi=0.3$ and $\Theta_1=\Theta_2=0$. Here only the anticrossings around $\varepsilon_D=U-V_Z$ can be seen. (d)-(f) show the differential conductance at $\varepsilon=0$ and $\phi=0.3$ for various spin canting angle configurations [(d) $\Theta_1=\Theta_2=0.8$, (e) $\Theta_1=0$, $\Theta_2=1.4$, (f) $\Theta_1=\Theta_2=0$]. All plots show a diamond-like lineshape. The dashed lines indicate line cuts shown in (g), (h) as a function of dot level energy. (g) corresponds to the upper row of plots with $eV=25\Gamma$ and (h) corresponds to the lower row of plots with $eV=22\Gamma$. In (g) and (h) the kets denote the spin ground state of the QD that changes when passing a Fano resonance.}
\label{didvdot}
\end{figure*}
The mean field approximation for the QD calls for a self-consistent treatment of the problem which we present later in Sec.~\ref{sec:numerical}~B. Here, we use an analytical approximation for illustration purposes. We use the approximation 
\begin{align}
	\braket{n_\uparrow}&=\vartheta(-U-V_Z-\varepsilon_D)\notag\\
	\braket{n_\downarrow}&=\vartheta(V_Z-\varepsilon_D),
	\label{density}
\end{align}
where $\vartheta(x)$ is the Heaviside function. This approximation corresponds to the ground state expectation value of an isolated QD. In order to have a smooth differential conductance we use the fact that the Heaviside function can be written as $\vartheta(x)=\lim_{n\to 0}\left(1/2+\arctan(x/n)/\pi\right)$.
Instead of performing this limit we use $n=10^{-4}$ for the following calculations.

Moreover, we use the parametrization
\begin{align}
	t_{1}=t\cos(\phi) \quad &t_{2}=t\sin(\phi)\notag\\
	t_{D1}=t_{\text{Dot}}\sin(\phi) \quad &t_{D2}=t_{\text{Dot}}\cos(\phi),
\end{align}
and define the tunneling width $\Gamma=2\pi\nu(0)|t|^2$. In this parametrization the single parameter $\phi$ controls the strength of nonlocal couplings where for $\phi=0$ only couplings to the nearest MBS exist, while for $\phi=\pi/4$ the coupling to both MBSs is identical. We choose this parametrization, because in this way the relative strength between nonlocal and local couplings is the same on both sides of the nanowire which we would expect because of its spatial symmetry.

The CGF at zero temperature has the same form as in Eq.~(\ref{cgfwd}) which means that only Andreev reflection in two different spin channels contribute to the electronic transport. The probability amplitudes for these Andreev processes have the property $p_\pm(V)=p_\pm(-V)$ which reflects the particle-hole symmetry of the superconductor hosting the MBSs. Because the spin-quantization axis of the QD is given by the orientation of the Zeeman field,  spin rotation invariance is lost and therefore both spin-canting angles enter the differential conductance independently in contrast to the scenario without QD where only their difference matters.

In general, the differential conductance has local maxima (resonances) and local minima (antiresonances). For a given set of parameters the differential conductance will have six maxima as a function of $V$.
They correspond to the eigenenergies of the system without the coupling to the lead.

As seen in Fig.~\ref{didvdot}, the differential conductance shows anticrossings because of the hybridization of the QD states and the MBSs. In general, at points in parameter space where the occupation number changes the differential conductance shows a discontinuity which we can attribute to the approximations we made using Eq.~(\ref{density}).
Away from the anticrossings the resonances can be attributed to either the dot states or the MBSs. The resonances corresponding to the dot states have a reduced width  at higher bias voltage because our model does not connect the dot with the lead directly. So all transport processes which contribute to the current need to include the low-energy MBSs.

The general form of the resonances resembles bowtielike (upper row of Fig.~\ref{didvdot}) or diamondlike (middle row of Fig.~\ref{didvdot}) patterns and can be used to analyze the Majorana nonlocality as discussed in Ref.~\onlinecite{Prada2017,Clarke2017}.

At fixed bias voltage between lead and superconductor Fano resonances can be found as a function of dot level energy $\varepsilon_D$ which is shown in Figs.~\ref{didvdot}(g) and \ref{didvdot}(h). These Fano resonances come in pairs (for electron- and holelike excitations on the QD). In the case of coupling to only one MBS these two Fano resonances are approximately symmetric with respect to each other, because a single MBS couples to electron and hole degrees of freedom in the same way. Mathematically, this can be explained if we examine the low-energy Hamiltonian. 
For a large Zeeman field and large Coulomb interaction $V_Z,U>\varepsilon$ and for dot level energies $\varepsilon_D$ where the occupation of the dot can change (i.e., if a spin level is close to zero energy) the low-energy physics is described by
\begin{align}
H_{LE}=&i\varepsilon\gamma_1\gamma_2+\varepsilon_{D2\uparrow(\downarrow)}d^\dagger_{\uparrow(\downarrow)} d_{\uparrow(\downarrow)}\notag\\
&+i\gamma_2\left[t_{D2\uparrow(\downarrow)}d^\dagger_{\uparrow(\downarrow)}+t_{D,\uparrow(\downarrow)}^*d_{\uparrow(\downarrow)}\right],
\end{align}
where we projected out the higher energy dot state, accordingly and $\varepsilon_{D,\sigma}=\varepsilon_D+\sigma V_Z + U \braket{n_{\overline{\sigma}}}$. This Hamiltonian is invariant under the transformation 
\begin{equation}
\varepsilon_{D,\uparrow(\downarrow)}\rightarrow-\varepsilon_{D,\uparrow(\downarrow)}\quad d_{\uparrow(\downarrow)}\rightarrow d^\dagger_{\uparrow(\downarrow)},
\end{equation}
up to a phase that can be gauged away. This reflects the particle-hole symmetry of an isolated MBS and is only present if the dot couples to a single MBS.

These Fano resonances arise, because there are basically two different transport paths in which Cooper pairs are transferred from the lead to the superconductor~\cite{Schuray2017}. In the first path the electrons from the lead enter the superconductor, virtually occupying the dot and then enter the condensate, whereas the second path is just the Andreev reflection where the two electrons from the lead directly enter the Cooper pair condensate. Here, the first path is resonant with respect to the dot level energy $\varepsilon_D$ while the second path is nonresonant. This results in an interference pattern known as Fano resonances.

\begin{figure*}
	\centering
\includegraphics[width=\textwidth]{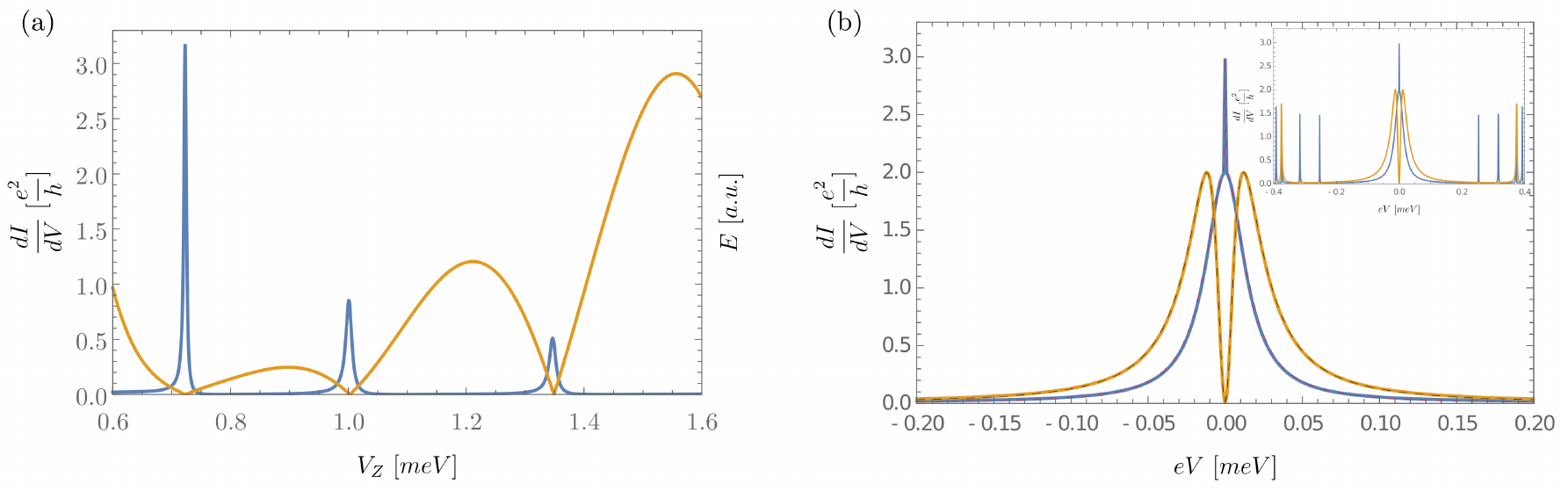}
\caption{(a) Zero bias differential conductance at zero temperature (blue) and energy $E$ of the lowest energy excitation in the nanowire (yellow) as a function of the applied Zeeman field for the setup without QD. The peaks in the differential conductance correspond to zero energy states in the nanowire. (b) Differential conductance vs. applied bias voltage energy at fixed Zeeman field $V_Z=0.7236$ meV (blue) and $V_Z=1.2$ meV (yellow). The Zeeman field values correspond to a small Majorana splitting (blue) and large splitting energy (yellow). The dashed lines are fits to the numerical data using Eq.~(\ref{didvwd}) with $\delta\Theta$, $\Gamma_1$, and $\Gamma_2$ as free parameters. The inset shows the differential conductance for a larger bias voltage energy window which includes higher energy states in the superconducting gap. The other microscopic parameters are $\mu=0$, $m^*=0.015m_e$, where $m_e$ is the electron rest mass, $\alpha=20$ meVnm, $L=1.1$ $\mu$m, and $\Delta=0.5$ meV.}
	\label{fig:numwithout}
\end{figure*}
In a bias window in which the anticrossings between the dot states and the MBSs exist no Fano resonances can be found. Due to the hybridization the states can no longer be identified as dot states or MBSs and thus the identification of two different paths is no longer possible. The resonances can also be linked to the spin states of the QD.
Because of this the Fano resonances can be symmetric even in the case of finite nonlocality ($\phi\neq0$) if only one of the MBSs can couple to the spin state of the QD. In general, the resonances can be found approximately at
\begin{equation}
	\varepsilon_{D,res}\approx\begin{cases}
		-U-V_Z\pm V & \text{for }\uparrow\text{ states on the dot},\\
		V_Z\pm V & \text{for }\downarrow\text{ states on the dot}. 
	\end{cases}
	\label{respos}
\end{equation}
For $V=0$, these resonances correspond to  changes of the occupation number in the ground state of the isolated dot. In the case of both MBSs spins pointing in the same direction along the quantization axis of the spins on the dot one spin state is decoupled from the nanowire and we recover the results for a spinless model which we already discussed in a previous work~\cite{Schuray2017}. Except for the discontinuity in the $dI/dV$ due to our approximations [Eq.~(\ref{density})], the only difference in this case is that the occupation number change in the ground state of the isolated dot is shifted in energy by the Zeeman field and the charging energy.

\section{Scattering matrix calculations for the full nanowire model}
\label{sec:numerical}
In this section we extend the effective low-energy model calculations by using the full model introduced in Eq.~(\ref{Hamiltonian}). We discretize the Hamiltonian on a chain in order to obtain a tight-binding Hamiltonian. To calculate the transport properties we use the python package Kwant~\cite{Groth2014} which utilizes the scattering formalism.
In this formalism the main entity is the scattering matrix
\begin{equation}
S=\left(\begin{matrix}S_{ee}&S_{eh}\\
S_{he}&S_{hh}\end{matrix}\right),
\end{equation}
where the blocks $S_{ij}$ connect the incoming modes of kind $i$ with outgoing modes of kind $j$ in the lead with $e$ and $h$ describing the electron and hole modes, respectively.
The differential conductance at zero temperature is
\begin{equation}
	\frac{dI}{dV}=\frac{e^2}{h}\left(N_e-T_{ee}+T_{eh}\right),
	\label{didvnum}
\end{equation}
where $N_e$ is the number of propagating electron modes in the lead~\cite{Blonder1982}. The transmission amplitudes $T_{ij}$, even though $T_{ee}$ is actually describing the reflection of an electron, can be calculated from the scattering matrix
\begin{equation}
	T_{ij}=\text{Tr}\left(S_{ij}^\dagger S_{ij}\right).
	\label{Transmission}
\end{equation}
\subsection{A. Setup without the dot}
First, we consider the Majorana nanowire without the dot attached and focus on short wires, in order to demonstrate the effects of a sufficiently large coupling to the second MBS.
In Fig.~\ref{fig:numwithout}, we present the differential conductance for a wire of length $L=1.1$ $\mu$m. In the topologically nontrivial regime ($V_Z>\sqrt{\Delta^2+\mu^2}$) the Majorana splitting energy oscillates as a function of applied Zeeman field. Whenever the splitting energy is vanishing there is a peak in the zero bias differential conductance [cf. Fig.~\ref{fig:numwithout}~(a)]. 
However, due to the short length of the wire the differential conductance is no longer quantized with $2e^2/h$. The height of the peak exceeds or undercuts $2e^2/h$ which can be attributed to the spin-canting angle differences as shown in the effective model calculations [see Eq.~(\ref{didvv0})].

The differential conductance as a function of bias voltage at fixed applied Zeeman field is shown in Fig.~\ref{fig:numwithout}(b). At a magnetic field which corresponds to a near zero Majorana splitting energy the differential conductance is again a sum of two Lorentzians with two different widths, however the smaller width is so small that the temperature of current state of the art experiments is too high to resolve them properly which has already been pointed out~\cite{Liu2017}.
At finite Majorana splitting the differential conductance does not deviate much from $2e^2/h$ at the resonances which can be attributed to the fact that the Majorana nonlocality is maximal at zero splitting~\cite{Penaranda2018}.

The low-energy transport in both cases can be very well described with the effective model calculations. This is shown by the fits (dashed lines) to the numerical data using Eq.~(\ref{didvwd}) with $\delta\Theta$, $\Gamma_1$, and $\Gamma_2$ as free parameters. What cannot be described with the effective model is the transport due to the higher energy states shown in the inset in Fig.~\ref{fig:numwithout}(b).

In recent experiments~\cite{Nichele2017} first hints of the nonlocal couplings could be seen, because the zero bias differential conductance peak exceeded $2e^2/h$ at low temperatures and large tunnel couplings. However, the full regime was not yet explored, because in these experiments the ratio of $k_BT/\Gamma$ was still too large to fully resolve a possible coupling to the more distant MBS.

\subsection{B. Setup with the dot}
\begin{figure*}
	\centering
\includegraphics[width=\textwidth]{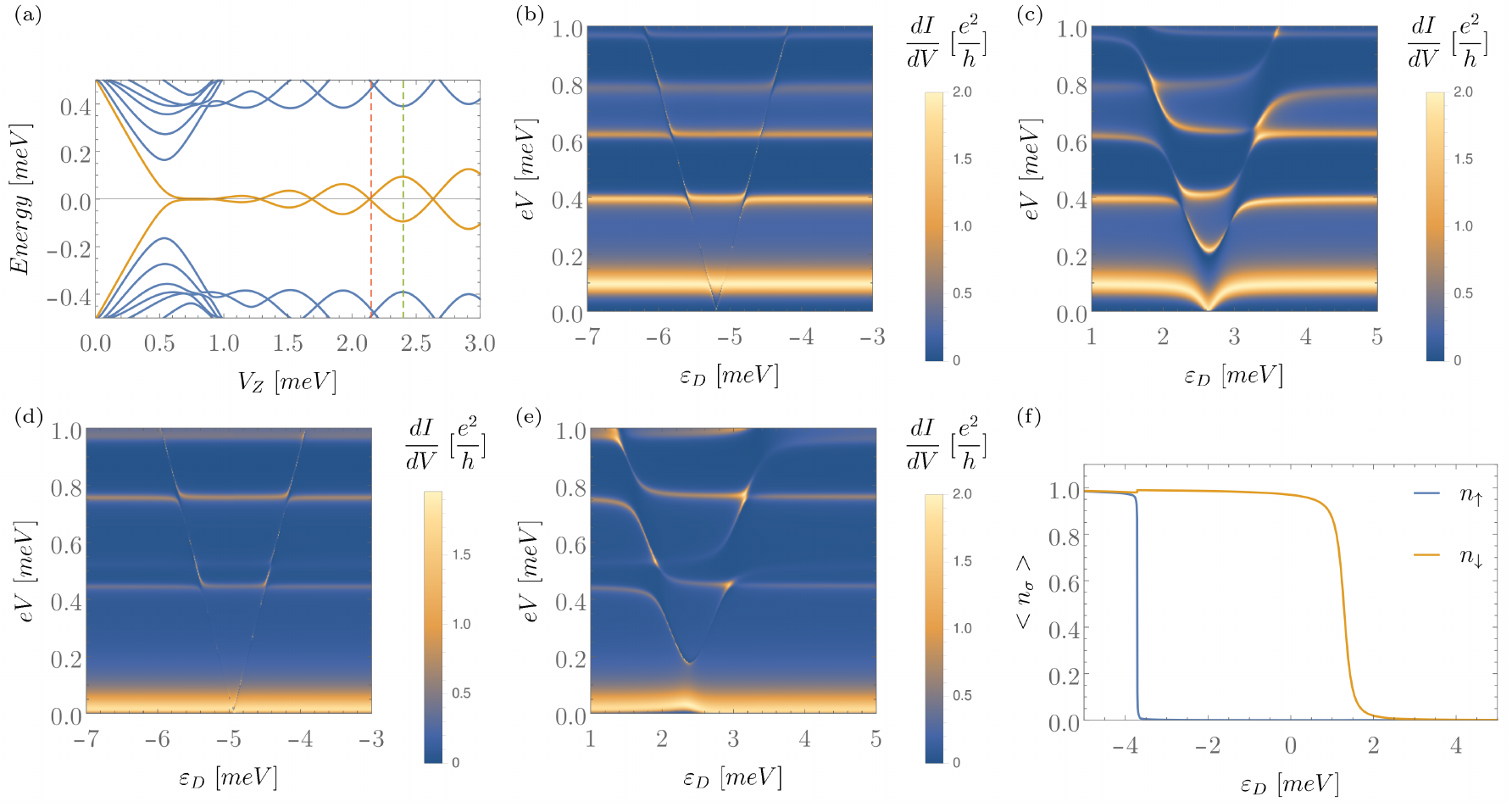}
\caption{(a) Spectrum of a finite-length nanowire-QD setup without an attached lead as a function of applied Zeeman field with $\varepsilon_D=-10$ meV. The vertical dashed lines indicate the Zeeman fields which were used in the calculations of the differential conductance in (b)-(e). (b)-(e) Differential conductance as a function of applied bias voltage between lead and Majorana nanowire and dot level energy for various Zeeman fields $V_Z=2.4$ meV [(b), (c)] and $V_Z=2.15$ meV [(d), (e)].  (f) Occupation number of the QD as a function of dot level energy for $V_Z=1.5$ meV. The other microscopic parameters are $\mu=0$, $m^*=0.015m_e$, where $m_e$ is the electron rest mass, $\alpha=20$ meVnm, $\Delta=0.5$ meV, $L=1.$ $\mu$m, and $U=3$ meV.}
	\label{fig:dotnum}
\end{figure*}
Because the nonlocal couplings have not yet been conclusively seen in experiments without a QD, we now focus on the setup containing the dot. In order to compare our numerical results with the analytical analysis we consider the weak coupling regime between dot and nanowire, i.e. the hopping between the dot and the nanowire is assumed to be only 10\% of the hopping inside the nanowire. Here, we use a self-consistent mean-field treatment of the QD in Eq.~(\ref{CoulombH}). The details of the algorithm that we used can be found in Appendix~\ref{sec:C}.
As seen in Fig.~\ref{fig:dotnum}(f) the occupation of the dot corresponds nearly to that of the isolated dot. The limitations of the self-consistent mean field theory is visible as a small discontinuity of the occupation number at the transition from a doubly to a singly occupied QD. Nevertheless, the self-consistent solution leads to much smoother transitions than the approximation made in Eq.~(\ref{density}) which only captures the transitions for the ground state of the isolated dot.

To compare the numerical analysis to our effective model calculations, we focus on the topologically nontrivial regime. As seen in Fig.~\ref{fig:dotnum}(a) this regime is dominated by the oscillating energy of the near zero energy Majorana states~\cite{DasSarma2012,Rainis2013,Prada2017}. However, not only the MBSs emerge inside the gap, but also other low-energy bound states can be found. The dot states hybridize with the MBSs as well as with the other states in the wire.
Figures~\ref{fig:dotnum}(b) and \ref{fig:dotnum}(c) show the differential conductance as a function of applied bias and dot level energy. The resonances reveal that there is a large splitting energy for the MBSs. However, the line shapes at low-energy are symmetric around the dot level energies where the occupation number of the dot changes.
This indicates, according to our effective model, that there is only an insignificant or even vanishing coupling to the second MBS.

The hybridization of the dot spin up state with the MBSs is much smaller than that of the dot's spin down state. This can be explained with the spin-canting angle of the MBSs. The different hybridizations can therefore also be used to analyze the spin-canting angle~\cite{Deng2018}.

In Figs.~\ref{fig:dotnum}(d) and \ref{fig:dotnum}(e) the Zeeman field is tuned in such a way that the Majorana splitting energy is close to zero. In the low-energy transport regime ($|eV|<0.3$ meV) the resonances are asymmetric around the point where the occupation number changes which indicates the existence of a large nonlocal coupling. This is consistent with our findings in the previous section for the setup without the QD.

The higher energy states also hybridize with the QD states.
This hybridization is also asymmetric for electron- and hole-like excitations on the dot. It comes from the fact that all excitations can be decomposed into two MBS components, and, in general, these components are not spatially separated and thus the dot couples to both Majorana components of each higher energy state.

The advantage of this setup compared to the setup without QD is that we can efficiently tune the spectrum of the system which can be probed by electron transport. This leads to qualitative features (symmetric vs. asymmetric hybridization) that allows us to discriminate between the case of coupling to only one or both MBSs. In contrast, the differential conductance peak in the setup without QD only changes its height but not its position when we include nonlocal couplings and its height change might be hard to detect experimentally due to thermal broadening of the conductance resonances.

\section{Conclusion}
In conclusion, we calculated and analyzed the cumulant generating function (CGF) for a spinful normal conducting lead - Majorana nanowire - QD setup. The CGF shows that the only process contributing to the low-energy transport through the junction is Andreev reflection in two channels corresponding to the two spin channels of the lead. We used this CGF to calculate the transport properties of the system --- average current and symmetrized zero frequency noise.

We described the low-energy sector of the Majorana nanowire in the topological nontrivial regime with the two MBSs emerging at the ends of the nanowire. To account for the finite length of the nanowire we not only included a coupling from the lead or the dot to the closest MBS, respectively, but also to the more distant one and included a finite energy splitting for the two MBSs. We also took into account the spin dependent tunneling amplitudes to account for the spin canting of the two MBSs at the corresponding interfaces which allows for analytical transport results that depend on characteristic and tunable properties of the MBSs. 

For the system without the QD we found that the coupling to the second MBS has a larger impact when the Majorana splitting energy is small. At zero bias and zero temperature the differential conductance becomes a Lorentzian as a function of the Majorana splitting energy where the width is governed by the product of both tunneling rates to the lead and its height is given by the sine of the spin-canting angle difference of the two MBS wave functions at the interface. Moreover, the temperature dependence of the differential conductance shows that the influence of the second MBS is only revealed at low temperatures smaller than the tunneling width of the coupling to the more distant MBS. Furthermore, we showed that the Fano factor loses its quantization at zero bias and zero temperature due to the coupling to the more distant MBS.

We treated the Coulomb interaction on the dot within a mean field approximation. The coupling to the QD leads to additional resonances and as a function of dot level energy at fixed bias voltage up to four Fano resonances emerge. These Fano resonances come in pairs situated around the points where the occupancy of the dot changes. These Fano resonances within each pair are mirror symmetric in the case of coupling to only one MBS. When a coupling to both MBSs is present in our calculation this symmetry is broken independently of the splitting energy of the MBSs. As a function of dot level energy and bias voltage the resonances of the differential conductance show a bow-tie or diamond like structure for zero or finite splitting energy, respectively.

To support our analytical low-energy findings, we discretized the full Hamiltonian and analyzed the differential conductance obtained by a numerical scattering matrix calculation. Our results confirm earlier predictions based on spectral properties that the nonlocal couplings to the two MBSs is largest when the splitting energy is smallest~\cite{Prada2017,Penaranda2018} and that the spin-canting angle difference changes as a function of applied magnetic field~\cite{Sticlet2012,Prada2017,Serina2018,Schuray2018,Milz2019}.

Our results provide vital information on how experimentally one could tell pairs of true MBSs appearing at the wire's ends apart from nontopological MBSs (for a recent review on this topic, see Ref.~\onlinecite{Prada2019}). In particular, our concrete analytical transport results with characteristic parameter dependences, e.g., on the canting-angle difference of the two MBSs, should give ample opportunities to compare experimental results to our model calculations by tuning experimental knobs like magnetic field or gate voltages. In addition, the Fano resonances predicted in our proposed setup (with the QD attached to the far end of the wire) can only appear due to nonlocal 
processes over distant MBSs sitting at the ends of the wire. 

\begin{acknowledgments}
We gratefully acknowledge the support of the Lower Saxony PhD-programme “Contacts in Nanosystems”, the Braunschweig International Graduate School of Metrology B-IGSM and the DFG Research Training Group 1952 Metrology for Complex Nanosystems.
\end{acknowledgments}
\appendix
\section{Full counting statistics for networks of MBSs including spin }
\subsection{Calculation of the CGF}
Here, we outline the derivation of the cumulant generating function (CGF) for a system of coupled MBSs including the spin degree of freedom of the charge carriers along the lines of Ref.~\onlinecite{Weithofer2014}.
\label{sec:A}
The moment generating function  is defined as
\begin{equation}
	\chi(\lambda)=\braket{e^{i\lambda Q}},
\end{equation}
where the counting field $\lambda$ is coupled to the transferred charge $Q=\int_0^\mathcal{T}dt I(t)$. We introduce the auxiliary Hamiltonian
\begin{equation}
	H'=H-\frac{1}{2}\lambda(t)I,
	\label{H'}
\end{equation}
with
\begin{equation}
	\lambda(t)=\begin{cases}
		\lambda & t \in [0,\mathcal{T}]\quad\&\quad t\in\mathcal{C}_{-}\\
		-\lambda & t \in [\mathcal{T},0]\quad\&\quad t\in\mathcal{C}_{+}\\
		0 & \text{else}

	\end{cases},
	\label{countingcontour}
\end{equation}
where $\mathcal{C}_{\mp}$ is the forward (backward) part of the Keldysh contour and $\mathcal{T}$ is the time during which the measurement is performed.
With this auxiliary Hamiltonian we can rewrite
\begin{align}
	\chi(\lambda)&=\braket{T_C\exp\left(-i\int_\mathcal{C}dt H'(t)\right)}\\
	&=\braket{T_C\left(1-i\int_\mathcal{C}dt\left[H-(\lambda(t)/2)I\right]+(\dots)\right)}\notag\\
	&=\braket{1+i\lambda\int_0^{\mathcal{T}}dt I+(\dots)}=\braket{e^{i\lambda Q}}.\notag
	\label{chi}
\end{align}
Also, the current operator is the total derivative of the number operator in the lead with respect to time
\begin{align}
	I&=- \frac{d}{dt}N=-\frac{d}{dt}\int dx\sum_\sigma c^\dagger_\sigma(x)c_\sigma(x)\\
	&=i\left[H,\int dx\sum_\sigma c^\dagger_{\sigma}(x)c_\sigma(x)\right]\notag\\
	&=i\left[H_T,\int dx\sum_\sigma c^\dagger_{\sigma}(x)\psi_\sigma(x)\right]\notag\\
	&=\sum_{n,\sigma}\gamma_n\left(t_{L\sigma n}c_{\sigma}^\dagger(0)-t^*_{L\sigma n}c_{\sigma}(0)\right).\notag
	\label{I}
\end{align}
Next we consider the time-dependent unitary transformation $U_\lambda=e^{i\frac{\lambda(t)}{2}N}$ and apply it to $H$
\begin{equation}
	H\rightarrow H_\lambda = U_\lambda H U_\lambda^\dagger - i U_\lambda \dot{U_\lambda}^\dagger.
\end{equation}
This leads to
\begin{equation}
 H_\lambda'=H_\lambda - \frac{1}{2}\lambda(t) I= U_\lambda H U_\lambda^\dagger.
\end{equation}
For the calculation of $U_\lambda H_T U_\lambda^\dag$ we first point out that $U_\lambda$ commutes with all Majorana operators, so all we need to consider is
\begin{align}
 U_\lambda c_\sigma(x) U_\lambda^\dag
&=
    e^{i \lambda(t) N/2} c_\sigma(x) e^{-i \lambda(t) N/2}  \\
&=
    c_\sigma(x) + \frac{i \lambda(t)}{2} [N, c_\sigma(x)] + (\dots) \notag \\
&=
    c_\sigma(x) - \frac{i \lambda(t)}{2} c_\sigma(x) + (\dots) \notag
\end{align}
Therefore, we find
\begin{align}
 H'_\lambda=&\ H_L + H_M' \\
 &+ \sum_{n,\sigma}  t_{Ln\sigma} e^{i \lambda(t)/2} c_\sigma^\dag(0)\gamma_n + t_{Ln\sigma}^*e^{-i \lambda(t)/2} \gamma_n c_\sigma(0)  \notag\\
 &=H_L + H_M' +  H_T^{\lambda},\notag
\end{align}
where $H'_M$ is defined in Eq.~(\ref{Hs}) and $t_{Ln\sigma}$ is defined for $n=1,2$ in Eq.~(\ref{tparams}) and $t_{Ln\sigma}=0$ for $n>2$.
So we find
\begin{equation}
 \chi(\lambda)=\braket{T_C\exp\left(-i\int_\mathcal{C}ds H_L+H_M+H^\lambda_T\right)}.
\label{chifin}
 \end{equation}
 We can formulate this moment generating function in the continuum notation as a functional integral. It is important to have in mind that Majorana fermions are described by real field operators and therefore we introduce real Grassmann variables $\hat \gamma$ in order
to calculate the functional integral, while we need two mutually independent complex Grassmann variables $\hat c$ and $\hat{\overline c}$ for the electrons in the lead.

The functional integral to calculate the moment generating function is then given as
\begin{equation}
\chi(\lambda)=\int\mathcal{D}[\hat\gamma,\hat c,\hat{\overline c}]e^{iS^{\lambda}[\hat\gamma,\hat c,\hat{\overline c}]},
\end{equation}
where $S^{\lambda}[\hat\gamma,\hat c,\hat{\overline c}]=S_M[\hat\gamma]+S_{T}^{\lambda}[\hat\gamma,\hat c,\hat{\overline c}]+S_L[\hat c,\hat{\overline c}]$ is the Keldysh action, containing the action for the Majorana fermions, the action of the lead, and the action describing the tunneling from the lead to the system of Majorana bound states.

These parts are given as
\begin{align}
S_M[\hat\gamma]=&\sum_{\alpha\beta}\int_{\mathcal{C}}\int_{\mathcal{C}}d sd s' \hat\gamma_\alpha(s)[D^{(0)}(s,s')]_{\alpha\beta}^{-1}\hat\gamma_\beta(s')\notag\\
S_{T}^{\lambda}[\hat\gamma,\hat c,\hat{\overline c}]=&\sum_{\alpha\sigma}\int_{\mathcal{C}}d s  \bigg[t_{L\alpha\sigma}e^{ \frac{i \lambda(s)}{2}}\hat{\overline c}_\sigma(0,s)\hat\gamma_\alpha(s)\notag\\
&+ t_{L\alpha\sigma}^*e^{- \frac{i \lambda(s)}{2}}\hat \gamma_\alpha(s)c_\sigma(0,s) \bigg]\\
S_L[\hat c,\hat{\overline c}]=&\sum_{\sigma}\int_{\mathcal{C}}\int_{\mathcal{C}}d s d s'\hat{\overline c}_\sigma(0,s)[G_\sigma(s,s')]^{-1}\hat c_\sigma(0,s'),\notag
\end{align}
where, $D^{0}(s,s')$ is the unperturbed Green's function for the Majorana bound states and $G_\sigma(s,s')=G_\sigma(x'=0,x=0,s',s)=-i\braket{T_{\mathcal{C}}\hat c_\sigma(0,s')\hat{\overline c}_\sigma(0,s)}$ is the boundary Green's function for the lead with spin $\sigma$. The position integral for $x\neq0$ for the lead has already been performed and is neglected because
the path integral is normalized in such a way that $\chi(0)=1$.
Now the moment generating function only contains Gaussian integrals. We can integrate over the lead degrees of freedom to find
\begin{equation}
\chi(\lambda)=\int\mathcal{D}[\hat\gamma]\exp\left(i\sum_{\alpha\beta}\int_\mathcal{C} d s d s'\hat\gamma_\alpha(s) [D^{\lambda}(s,s')]^{-1}_{\alpha\beta}\hat\gamma_\beta(s')\right),
\end{equation}
where $[D^{\lambda}(s,s')]^{-1}=[D^{(0)}(s,s')]^{-1}-\Sigma^{\lambda}(s,s')$, with the counting field dependent self energy
\begin{align}
\Sigma^{\lambda}_{\alpha\beta}(s,s')=&\sum_\sigma\bigg[-t_{L\alpha\sigma}t^*_{L\beta\sigma}e^{-i\frac{\lambda(s)-\lambda(s')}{2}}G_\sigma(s,s')\\
&+t_{L\beta\sigma}t^*_{L\alpha\sigma}e^{i\frac{\lambda(s)-\lambda(s')}{2}}G_\sigma(s',s)\bigg].\notag
\end{align} 
We can now use the Gaussian integral for real valued Grassmann fields~\cite{Swanson2014} to find
\begin{equation}
\chi(\lambda)=\frac{\sqrt{\det\left([D^{\lambda}]^{-1}\right)}}{\sqrt{\det\left([D^{\lambda=0}]^{-1}\right)}},
\end{equation} 
where we enforced the normalization by division with $\sqrt{\det\left([D^{\lambda=0}]^{-1}\right)}$. The determinant has to be calculated with respect to time, Majorana and Keldysh indices.
During the long measuring time $\mathcal{T}$ the counting fields are constant and a Fourier transform diagonalizes the Keldysh Green's function in energy space, so that the determinant with respect to the energy space is just a product and therefore the cumulant generating function, the logarithm of $\chi$, is given by
\begin{equation}
\ln\chi(\lambda)=\frac{1}{2}\sum_\omega\ln\left[\frac{\det\left([D^{\lambda}]^{-1}(\omega)\right)}{\det\left([D^{\lambda=0}]^{-1}(\omega)\right)}\right].
\label{cumsum}
\end{equation}
Now the determinant has to be taken with respect to Keldysh and Majorana indices.
The summation can be transformed into an integration, because the frequencies will be quantized due to the long measuring time which results in the Levitov-Lesovik formula~\cite{Nazarov2003,Levitov2004,Weithofer2014}
\begin{equation}
\ln\chi(\lambda)=\frac{\mathcal{T}}{2}\int\frac{d\omega}{2\pi}\ln\left[\frac{\det\left([D^{\lambda}]^{-1}(\omega)\right)}{\det\left([D^{\lambda=0}]^{-1}(\omega)\right)}\right].
\label{LevitovII}
\end{equation}
\begin{figure*}
	\centering
\includegraphics[width=0.75\textwidth]{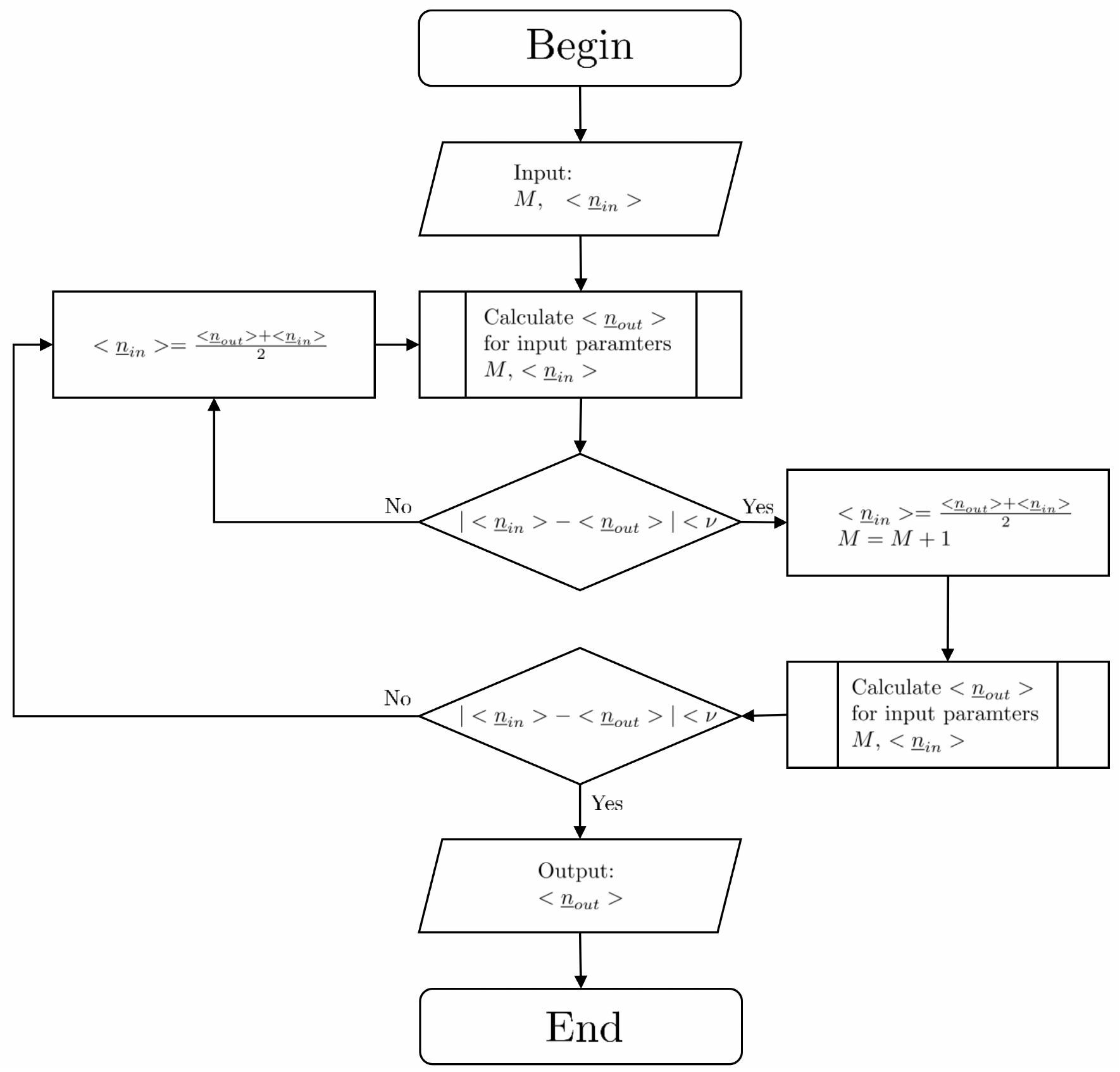}
	\caption{Program flow chart for the self-consistent calculation of the QD occupation number. The program contains two loops. In the first loop the the QD occupation number $\braket{\underline{n}}$ is calculated self-consistently for a fixed number of sites of the lead $M$, while the second loop increases the number of sites for the lead until convergence (with small convergence parameter $\nu$).}
	\label{fig:flowchart}
\end{figure*}
\subsection{Majorana Green's Function}
\label{sec:B}
In this section we want to describe the calculation of the Majorana Green's function in detail. We start with the Heisenberg equation of motion (EOM) for the Majorana operators with the unperturbed Hamiltonian $H_M'$ [see Eq.~(\ref{Hs})]
\begin{equation}
\frac{d}{dt}\gamma_\alpha=i[H_M',\gamma_\alpha]=2\sum_\beta A_{\alpha\beta}\gamma_\beta,
\end{equation}
where we used the skew symmetric property $A_{\alpha\beta}=-A_{\beta\alpha}$. The solutions to this EOM are
\begin{equation}
 \gamma_\alpha(t)=\sum_\beta\mathcal{B}_{\alpha\beta}(t)\gamma_\beta(0),
\end{equation}
with $\mathcal{B}(t)=\exp(2A t)$. The time dependent unperturbed Majorana Green's function then is
\begin{align}
 D^{(0)}_{\alpha\beta}(t)=&-i\braket{T_\mathcal{C}\gamma_\alpha(t)\gamma_\beta(0)}\\
 =&-i\mathcal{B}_{\alpha\beta}(t)\left(\Theta_\mathcal{C}(t)-\Theta_\mathcal{C}(-t)\right)-i\sum_{\nu\neq\beta}\mathcal{B}_{\alpha\nu}(t)e_{\nu\beta}\notag\\
 =&-i\mathcal{B}_{\alpha\beta}(t)\left(\begin{matrix}
                                        \text{sign}(t)&-1\\
                                        1&-\text{sign}(t)
                                       \end{matrix}
\right)\label{eq:unpert}\\&-i\sum_{\nu\neq\beta}\mathcal{B}_{\alpha\nu}(t)e_{\nu\beta}\left(\begin{matrix}
                                        1&1\\
                                        1&1
                                       \end{matrix}
\right)\notag,
\end{align}
where $\Theta_\mathcal{C}(t)$ is the Heaviside function on the Keldysh contour and we defined $e_{\nu\beta}=\braket{\gamma_\nu\gamma_\beta}$. Here, the Keldysh indices are organized as $[(--,-+),(+-,++)]^T$. In order to calculate the Fourier transform of $D^{(0)}_{\alpha\beta}(t)$ we need to consider the Fourier transforms of $\mathcal{B}(t)$ and $\mathcal{B}(t)\text{sign}(t)$. First, we note that we can diagonalize the Hermitian matrix $-iA=UD_AU^\dagger$ where $D_A=\text{diag}(\lambda_k)$ with the eigenvalues $\lambda_k$ where for every positive $\lambda_k$ there is a $\lambda_{\overline{k}}=-\lambda_k$ which in our case leads to six $\lambda_k$ and find
\begin{align}
 \int dt e^{i\omega t}\mathcal{B}(t)&=U\int dt e^{i\omega t} e^{iD_A t} U^\dagger\notag\\&=U\diag{2\pi\delta(\omega-\lambda_k)}U^\dagger.
 \label{firstfourier}
\end{align}
In order to calculate the second Fourier integral we find it convenient to use the following change of basis $2A=QSQ^T$ with
\begin{equation}
 S=\left(\begin{matrix}
    0&\lambda_1&0&0&0&0\\
    -\lambda_1&0&0&0&0&0\\
    0&0&0&\lambda_2&0&0\\
   0&0&-\lambda_2&0&0&0\\
   0&0&0&0&0&\lambda_3\\
   0&0&0&0&-\lambda_3&0
   \end{matrix}\right),
\end{equation}
which is possible because $A$ is skew symmetric.
Using this we find
\begin{align}
\label{secondfourier}
 &-i\int dt e^{i\omega t}\text{sign}(t)\mathcal{B}(t)=Q\int dt e^{i\omega t}e^{St} Q^T\\
 &=-iQ\int dt \text{sign}(t)e^{i\omega t}\text{diag}\left(\begin{matrix}
            \cos\lambda_k t&\sin\lambda_k t\\
            -\sin\lambda_k t&\cos\lambda_k t
                                          \end{matrix}
\right) Q^T\notag\\
&=Q\text{diag}\left(\begin{matrix}
            \frac{2\omega}{\omega^2-\lambda_k^2}&\frac{-2i\lambda_k}{\omega^2-\lambda_k^2}\\
            \frac{2i\lambda_k}{\omega^2-\lambda_k^2}&\frac{2\omega}{\omega^2-\lambda_k^2}
                                          \end{matrix}
\right) Q^T\notag.
\end{align}
For $\omega\neq\lambda_k$ the off diagonal blocks and all terms containing $e_{\nu\beta}$ in Eq.~(\ref{eq:unpert}) vanish so that $D^{(0)}_{\alpha\beta}(\omega)$ is block diagonal. For $\omega=\lambda_k$ the Dirac distribution in Eq.~(\ref{firstfourier})  as well as the terms in Eq.~(\ref{secondfourier}) diverge so that the inverse of the corresponding $\lambda_k$ block of the Green's function vanishes. It would vanish even if we would neglect the terms coming from Eq.~(\ref{firstfourier}).
So we only need to consider the block diagonal part of $D^{(0)}_{\alpha\beta}(\omega)$ for its inverse and find
\begin{equation}
 [D^{(0)}]^{-1}=\left(\begin{matrix}
                       [D^{(0)--}]^{-1}&0\\
                       0&-[D^{(0)--}]^{-1}
                      \end{matrix}
\right),
\end{equation}
with
\begin{align}
 [D^{(0)--}]^{-1}=&\ Q\text{diag}\left(\begin{matrix}
                \frac{\omega}{2}&\frac{i\lambda_k}{2}\\
                -\frac{i\lambda_k}{2}&\frac{\omega}{2}
                                     \end{matrix}
\right)Q^T\\
 =&iA+\frac{\omega}{2}I.\notag
\end{align}

Following Ref.~\onlinecite{Kamenev2011} the lead boundary Green's function $ G_\sigma(t',t)=G_\sigma(x'=0,x=0,t',t)=-i\braket{T_{\mathcal{C}}c_\sigma(x'=0,t')c_\sigma^\dagger(x=0,t)}$ for spin $\sigma$ can be written in the Keldysh-rotated basis in matrix form as
\begin{equation}
 iG_\sigma(t',t)=\pi\nu(0)\left(\begin{matrix}
                                 \delta(t-t')&2F_\sigma(t-t')\\
                                 0&-\delta(t-t')
                                \end{matrix}
\right),
\end{equation}
where $\nu(0)$ is the density of states per spin at the Fermi level in the lead and the Fourier transform of the distribution matrix is $F_\sigma(\omega)=1-2n_\sigma(\omega)$ with the Fermi distribution function $n_\sigma(\omega)=(1+e^{\omega/k_B T})^{-1}$.

After a back rotation and Fourier transform we find
\begin{equation}
 G_\sigma(\omega)=i2\pi\nu(0)\left(\begin{matrix}
                                 n_\sigma(\omega)-\frac{1}{2}&n_\sigma(\omega)\\
                                 n_\sigma(\omega)-1& n_\sigma(\omega)-\frac{1}{2}
                                \end{matrix}
\right).
\end{equation}
\section{Self-consistent algorithm}
\label{sec:C}

To incorporate the lead in the self-consistent calculation for the QD occupation number we consider it to have a finite length with $M$ sites. We also start with a given input vector for the spin up and spin down occupation numbers $\underline{n}_{in}$ on the dot and calculate the expectation value in the ground state of Eq.~(\ref{Hfull}) for the number operator on the dot $\underline{n}_{out}$. If the difference between input and output $|\underline{n}_{in}-\underline{n}_{out}|$ is larger than a predefined value $\nu$ (in our case $\nu=0.001$) we use $\underline{n}_{in}=0.5(\underline{n}_{in}+\underline{n}_{out})$ as the new input value and start the calculation again. Once the difference between input and output is smaller than $\nu$, we add a site to the finite size lead. Then we use $\underline{n}_{in}=\underline{n}_{out}$ and calculate the expectation value in the ground state for the number operator on the dot again. If the output after adding a site does not change more than $\nu$ we consider the system to have converged and use the output as the occupation number for the calculation of the differential conductance. If it is larger than $\nu$ we start the program again with the lead being one site larger than before. The program flow chart for this method is shown in Fig.~\ref{fig:flowchart}.

\end{document}